\documentclass[%
 aip,
 amsmath,amssymb,
 reprint,%
]{revtex4-1}

\usepackage{graphicx}
\usepackage{dcolumn}
\usepackage{bm}

\usepackage[utf8]{inputenc}
\usepackage[T1]{fontenc}
\usepackage{mathptmx}
\usepackage{etoolbox}

\usepackage{amsmath, mathtools, subcaption, tabularx}
\usepackage[noabbrev,capitalise,nosort,nameinlink]{cleveref}
\usepackage[inkscapeformat=png]{svg}
\usepackage{enumitem, moreenum}
\setlist[enumerate]{nosep}
\usepackage[mathscr]{euscript}

\newcommand*{\defeq}{\coloneqq}

\newcommand*{\Reals}{\mathbb{R}}

\makeatletter
\def\@email#1#2{%
 \endgroup
 \patchcmd{\titleblock@produce}
  {\frontmatter@RRAPformat}
  {\frontmatter@RRAPformat{\produce@RRAP{*#1\href{mailto:#2}{#2}}}\frontmatter@RRAPformat}
  {}{}
}%
\makeatother
\begin{document}

\title{Bayesian optimisation of poloidal field coil positions in tokamaks}
\author{Timothy Nunn}
\email{timothy.nunn@ukaea.uk}
\author{Kamran Pentland}%
\author{Vignesh Gopakumar}
\author{James Buchanan}
\affiliation{ 
UKAEA (United Kingdom Atomic Energy Authority), Culham Campus, Abingdon, Oxfordshire, OX14 3DB, United Kingdom
}%

\date{\today}

\begin{abstract}
The tokamak is a world-leading concept for producing sustainable energy via magnetically-confined nuclear fusion.
Identifying where to position the magnets within a tokamak, specifically the poloidal field (PF) coils, is a design problem which requires balancing a number of competing economical, physical, and engineering objectives and constraints. 
In this paper, we show that multi-objective Bayesian optimisation (BO), an iterative optimisation technique utilising probabilistic machine learning models, can effectively explore this complex design space and return several optimal PF coilsets.
These solutions span the Pareto front, a subset of the objective space that optimally satisfies the specified objective functions.
We outline an easy-to-use BO framework and demonstrate that it outperforms alternative optimisation techniques while using significantly fewer computational resources.
Our results show that BO is a promising technique for fusion design problems that rely on computationally demanding high-fidelity simulations.
\end{abstract}

\maketitle

\section{Introduction} \label{sec:intro}

\subsection{Motivation and aims}

A spherical tokamak is a torus-shaped device with a low aspect ratio that uses strong magnetic fields to confine and control a thermonuclear fusion plasma, with the goal of producing fusion energy \citep{freidberg2008plasma}.
The Spherical Tokamak for Energy Production (STEP), currently in the design phase and targeting completion in 2040 \citep{step}, is one of a few ongoing fusion research and development projects based on the spherical tokamak concept.
To deliver fusion power to the grid on such a short timescale, researchers are increasingly designing next-generation tokamaks in silico with multi-physics simulations, many of which require high-performance computing (HPC) resources.

STEP is no exception, with initial concept designs for the tokamak (and the associated plasma) being generated by low-fidelity integrated modelling codes \citep{muldrew2024conceptual} such as PROCESS \citep{Morris_PROCESS_2024, muldrew2020process} and Bluemira \citep{bluemira, coleman2019blueprint, franza2022mira}.
These codes use simplified physics and engineering models to produce designs within seconds or minutes. 
In contrast, more complex medium- to high-fidelity codes, such as JINTRAC \citep{romanelli2014jintrac}, incorporate more detailed physics models but can require days or weeks to complete a single simulation. 
These higher-fidelity simulations play a crucial role in refining, integrating, and validating the initial concept design across the entire fusion power plant \citep{step_digital}.
Making the most efficient use of these computationally expensive simulations is critical if we wish to accelerate the design of future fusion power plants like STEP.

Our focus here will be on the design of the poloidal field (PF) coil system, which plays a critical role in controlling the position and shape of the plasma in both the core and divertor regions of the tokamak \citep{lim2010design}.
In particular, some coils are crucial for managing the vertical stability of elongated plasmas, such as those in spherical tokamaks, where the higher elongation can lead to larger vertical instability, risking disruption without appropriate control \citep{anand2023modelling}.
By generating poloidal magnetic fields, the PF coil system ensures the plasma remains in equilibrium, balancing the inward-facing magnetic forces produced by the coils against the outward-facing pressure-driven forces generated by the plasma \citep{wesson2011tokamaks}.
The design of the system---in terms of the coil positions, sizes, and shapes---will have a significant impact on plasma performance and stability and will therefore need to satisfy a number of competing (and often conflicting) constraints.
In the plasma, for example, constraints are required to ensure X-points form in specific locations (for stability), strikepoints hit the correct divertor plates (for heat management), and total current density limits on the PF coils are not exceeded.
In terms of the tokamak itself, the locations/sizes of the coils will inevitably be constrained by the vacuum vessel, diagnostic systems, and maintenance ports (to name but a few).

In addition to constraints, there will be a number of objectives related to the desired operational plasma conditions that we wish the chosen coilset to minimise or maximise (depending on the objective).
This could include minimising the coil size to reduce fabrication, construction, and installation costs or could include minimising current flows to reduce power consumption and structural stresses from forces produced by the coils \citep{coleman2020design}.
Moreover, we may wish to optimise certain properties of the plasma in the divertor chambers in order to minimise heat loads on plasma facing components and improve exhaust performance \citep{hudoba2023divertor,hudoba2024}.
Simultaneously satisfying both the objectives and constraints will require the solution of a complex optimisation problem that needs to be tackled in a systematic, computationally efficient manner. 

In this paper, we will perform multi-objective Bayesian optimisation (BO) on an earlier baseline design of the STEP PF coil system \citep{hudoba2023magnetic}.
Our aims are to:
\begin{enumerate}[label=(\roman*)]
    \item design and outline an easy-to-use BO framework which is flexible, data efficient (reducing the computational cost of design), and can yield more optimal designs than obtained through other exhaustive optimisation schemes. 
    \item identify a \emph{Pareto front}, i.e. a set of optimal PF coil locations, that outperform the baseline for some given objectives and constraints.
    \item motivate more widespread adoption of BO for the in silico design of interlinked components on future tokamak devices to save time, minimise financial costs, and improve plasma performance. 
\end{enumerate}

We should stress that this work has not had a direct impact on the current design of the STEP PF coil system \citep{nasr2024} and is instead a demonstration of a generalisable BO framework for PF coil system design.
We do wish to highlight, however, that the framework is completely machine agnostic and can be used with different objectives and constraints to the ones we use here. 
It is the hope that frameworks such as this will be adopted more regularly within the integrated modelling codes currently used for tokamak design.

\subsection{Related work}

PF coilsets are typically optimised using integrated modelling codes for tokamak power plant design.
A common approach is to force the PF coils to lie on a contour ``rail’’ that surrounds the core plasma, reducing the number of degrees of freedom in the optimisation problem \citep{meneghini2024, coleman2020design}. 
Exclusion zones along the rails enforce engineering constraints, before nonlinear (non-Bayesian) optimisation is performed with respect to some pre-specified objectives and constraints on the plasma boundary shape.

While well-established, rail-based methods can restrict the PF coil design space, often rely on estimated objective function gradients, and can struggle with multiple competing objectives.
They are primarily suited to conventional aspect ratio tokamaks, where PF coil rails are placed outside (and close to) the toroidal field (TF) coils, sometimes leading to intersection issues.
BO, on the other hand, performs gradient-free global optimisation, can handle diverse constraints, and uses a surrogate model of the multi-output objective function to intelligently guide function evaluations.
This helps balance exploration of new designs and exploitation of known optimal designs, leading to high levels of data efficiency.

Despite these advantages, the adoption and application of BO in fusion engineering and design has, so far, remained relatively limited.
\citet{brown_multi-objective_2024}\citep{brown_sample-efficient_2024} aimed to improve six key properties of the safety factor profile by using BO on the current profiles in STEP.
They also demonstrate that BO performs better than a genetic algorithm with the same number of black-box function evaluations (as we will do later on).
\citet{mehta2024} use BO to find the parameters such as neutral beam injection power, plasma current, and plasma elongation in the DIII-D tokamak that safeguard against disruption during the ramp-down phase.
Similarly, \citet{pusztai2023bayesian} use BO to mitigate the impact of disruptions in ITER by exploring how injected deuterium and neon can minimise runaway electron currents, transported heat, and quench time post-disruption. 
\citet{jarvinen2022bayesian} also investigate runaway electron currents using BO as an advanced sampling method to help calibrate uncertainty and minimise the discrepancy between simulations and experimental data.
For fusion component design, \citet{humphrey2023} demonstrate the use of BO to minimise stresses in parametrised divertor monoblocks under fusion conditions.
The most relevant work to ours is that of \citet{nunn2023}, who use multi-objective BO to optimise TF coil shapes to reduce both financial costs and magnetic ripples (which affect plasma stability and performance).
In contrast, our approach deals with more computationally expensive, failure-prone plasma equilibrium simulations without analytic objective/constraint functions, necessitating the use of a classifier alongside the surrogate model.

The work here is inspired by that of \citet{hudoba2023magnetic}, in which the authors seek to optimise the STEP PF coil system by minimising deviations of key plasma parameters from a baseline scenario (which we adopt) and coil currents, while maximising divertor performance metrics.
Using a free-boundary equilibrium solver, thousands of potential PF coilsets are sampled and evaluated (in a Monte Carlo-type approach) before optimal solution sets are identified heuristically.
We aim to provide and fully outline an alternative, much more data efficient, framework for carrying out similar multi-objective optimisation that can return a STEP equilibrium similar to the baseline.

There are also a number of areas in fusion design where BO has yet to be applied but could potentially offer significant benefits. 
For example, parameter scans for optimal magnetic sensor placement, as explored for TCV \citep{romero2013} and SPARC \citep{stewart2023}, could benefit from BO's sample efficiency, saving computational resources and time. 
Similarly, these benefits could transfer to existing frameworks for stellarator coil design \citep{jorge2024, giuliani2024, kaptanoglu2025}. 

\subsection{Outline}
In \cref{sec:bayesopt}, we describe the multi-objective BO problem, the Gaussian process surrogate model, the classifier scheme, and the acquisition function required in the BO loop.
We follow this in \cref{sec:design-problem} by defining the PF coil design problem in terms of the input space, the objectives we seek to optimise, and the constraints on the plasma and the machine. 
In addition, we describe the simulator used to generate the plasma equilibria for each PF coilset and define cases in which the simulator may fail to produce a valid equilibrium (requiring the classifier).  
The numerical experiments are detailed and presented in \cref{sec:results}.
To highlight the data efficiency of the BO scheme, we carry out a number of experiments with a fixed computational budget and assess performance against alternative optimisation methods. 
In \cref{sec:conclusion}, we discuss our findings, highlight any major advantages and disadvantages of the BO framework applied to this problem, and propose avenues for future work. 

\section{Multi-objective Bayesian optimisation} \label{sec:bayesopt}

BO is a method for performing gradient-free global optimisation of black-box functions, typically utilised when the function is expensive-to-evaluate \citep{garnett_bayesoptbook_2023}. 
Practitioners will often want to identify (feasible) optimal points of the function's input/output spaces with as few function evaluations as possible---especially if there is a limited computational budget. 

Here, we are interested in optimising the nonlinear function $\bm{f}\colon \mathscr{U}\subseteq\Reals^d \to \Reals^{l+m}$ that takes in a $d$-dimensional input and returns $l$ objectives and $m$ constraints.
More formally, the aim of \emph{multi-objective BO} ($l>1$) is to solve 
\begin{equation} \label{eq:optim_problem}
    \operatorname*{argmin}_{\substack{\bm{x}\in\mathscr{U} \\ \bm{f}_{[l+1..l+m]}\preccurlyeq0}} \bm{f}_{[1..l]} \left( \bm{x} \right),
\end{equation}
where $\bm{f}_{1..k}$ denotes the first $k$ components of $\bm{f}$ and $\preccurlyeq$ denotes a component-wise less than or equal to comparison.

Given we need to optimise over multiple competing objectives, problems such as \eqref{eq:optim_problem} will often involve trade-offs where improving one objective may come at the expense of another.
The aim is therefore to seek the set of \emph{Pareto optimal} solutions $\mathscr{P}$ that are not \emph{dominated} by any other solutions. 
A solution $\bm{x}$ \emph{dominates} another $\bm{x}'$, denoted $\bm{x} \prec \bm{x}'$, if and only if $\bm{f} (\bm{x}) \preccurlyeq \bm{f}(\bm{x}')$ and $\exists j \in \{1,\ldots,l \}$ such that $\bm{f}_j(\bm{x}) < \bm{f}_j(\bm{x}')$.
In short, a solution $\bm{x}$ dominates $\bm{x}'$ if it is at least as good in all objectives and strictly better in at least one.
Given a dataset
\begin{equation*}
   \mathscr{D} = \left\{ \left( \bm{x}_i, \bm{f}(\bm{x}_i) \right) \right\}_{i=1}^N,
\end{equation*}
consisting of $N$ evaluations of $\bm{f}$, the \emph{Pareto set} for \eqref{eq:optim_problem} is defined as
\begin{align}
    \mathscr{P}(\mathscr{D}) \defeq \{ \bm{x} \in \mathscr{D} \mid \nexists \ \bm{x'} \in \mathscr{D} \ \text{s.t.} \ \bm{x}' \prec \bm{x} \}.
\end{align}
The \emph{Pareto front}, denoted $\mathscr{P}_{\bm{f}}$, is defined as the image of the Pareto set, i.e. $\mathscr{P}_{\bm{f}} \defeq \{ \bm{f}(\bm{x}) \mid \bm{x} \in \mathscr{P}(\mathscr{D}) \}$.
See \citet[Chp. 11.7]{garnett_bayesoptbook_2023} for an illustration of the Pareto front.

\subsection{The Bayesian optimisation loop}

The key component in BO for identifying feasible and optimal trade-offs between the objectives is a \emph{probabilistic surrogate model}, capable of performing uncertainty based exploration.
This model is typically trained on some initial dataset by maximising its marginal likelihood---more details on this surrogate model are given in \cref{sec:gps}.

\begin{figure}[t!]
    \centering
    \includesvg[width=0.47\textwidth, keepaspectratio]{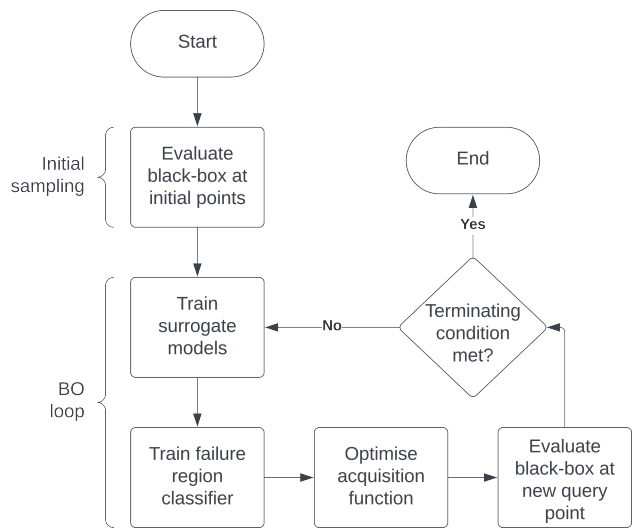}
    \caption{A flowchart illustrating the standard ``BO loop'' along with an additional failure region classifier step---see text for more details.}
    \label{fig:bo-flowchart}
\end{figure}

The first stage in BO (refer to \cref{fig:bo-flowchart}) is to construct this initial dataset (which we will call $\mathscr{D}$) by taking $N$ samples $\bm{x} \in \mathscr{U}$ and evaluating them all using $\bm{f}$. 
One popular method used is Sobol sampling \citep{sobol1967distribution}, whereby samples are chosen quasi-randomly with low discrepancy to achieve approximately uniform coverage of the input space.
The number of samples $N$ chosen/required may depend on the size of $d$, the computational budget available, and if parallel processing is available (for the $\bm{f}$ evaluations). 
Note that at this point, while we could use $\mathscr{D}$ to immediately generate a Pareto set $\mathscr{P}(\mathscr{D})$, this would almost certainly be a poor estimate given a lack of data points and that most would reside in non-optimal regions of the objective space.

It is worth noting that for many black-box functions such as $\bm{f}$, there will be \emph{failure regions} of the input space that cannot be evaluated.
The reasons for failure in our particular setting are discussed further in \cref{sec:failure}.
During construction of the initial dataset, samples that lie in failure regions may be encountered and so we do not wish to include these in the dataset. 
We do, however, wish to learn from these samples so that we do not encounter similar samples again and therefore we train a \emph{classifier} to predict when this may happen---a similar approach was taken by \citet{hornsby2024gaussian} when generating gyrokinetic simulation datasets for spherical tokamaks.
This classifier will be used within the \emph{BO loop}, which can be seen in \cref{fig:bo-flowchart} and is now outlined:
\begin{description}
    \item[Stage 1] Generate the initial dataset $\mathscr{D}$ using the Sobol sampling scheme. 
    \item[Stage 2] Train the surrogate model using the dataset $\mathscr{D}$ (see \cref{sec:gps}) to obtain a probabilistic model: $\bm{f}(\bm{x}) \mid \mathscr{D}$. 
    Note that we exclude \emph{failure} samples during training as they do not provide any valid information about the objective or constraint values.
    \item[Stage 3] Train the failure region classifier using the generated data (see \cref{sec:classifier}). 
    \item[Stage 4] Maximise the \emph{acquisition function} over the input space to identify the most ``informative'' point $\bm{x}^*$ to observe next, given the current surrogate model (and classifier) $\bm{f}(\bm{x}^*) \mid \mathscr{D}$ (see \cref{sec:acqf}).
    \item[Stage 5] Evaluate $\bm{f}(\bm{x}^*)$ and add it to the dataset: $\mathscr{D} \defeq \mathscr{D} \cup \{ (\bm{x}^*, \bm{f}(\bm{x}^*)) \}$.
    \item[Stage 6] Check whether the terminating condition is met and if not, return to Stage 2. 
    In our case, we check whether or not the maximum number of iterations has been exceeded (to remain within our computational budget). Other terminating conditions include stopping once improvements in the acquisition function are below some threshold or when the objectives are deemed to be sufficiently optimal \citep{ishibashi2023stopping}.
\end{description}

\subsection{The Gaussian process surrogate} \label{sec:gps}

In BO, the most common type of surrogate used is a Gaussian process (GP), which is a probabilistic machine learning model for performing inference on the value of a function $f \colon \mathscr{U} \to \Reals$ given some training data---see \citet[Chp. 2]{garnett_bayesoptbook_2023}.
It is characterised by a mean function $m \colon \mathscr{U} \to \Reals$ and a positive semi-definite covariance function $k \colon \mathscr{U} \times \mathscr{U} \to \Reals$ (with $k(\cdot,\cdot) \geq 0$) such that the prior can be defined as
\begin{equation} \label{eq:prior}
    f \sim \mathscr{GP} \left( m, k \right).
\end{equation}
The distribution of this prior is the joint distribution of (infinitely) many Gaussian random variables and can be thought of as a distribution over functions.
Therefore, at a finite set of evaluation points $\bm{X} = \{ \bm{x}_1, \bm{x}_2,\ldots \} \subset  \mathscr{U}$ we have that
\begin{align}
    f(\bm{X})\sim\mathscr{N}\left(\bm{\mu},\bm{\Sigma}\right),
\end{align}
where $\bm{\mu} = [m(\bm{x}_1), m(\bm{x}_2), \ldots]^{\intercal}$ is the mean vector and $[\bm{\Sigma}]_{i,j} = k(\bm{x}_i,\bm{x}_j) \ \forall i,j \in \{1,2,\ldots \} $ is the covariance matrix. 

Training a GP requires conditioning the prior \eqref{eq:prior} on the dataset of known function evaluations $\mathscr{D}$ (with outputs standardised to mean $0$, standard deviation $1$ and the inputs transformed to the unit hypercube) such that we obtain the following posterior distribution
\begin{align} \label{eq:posterior}
    f(\bm{X}) \mid \mathscr{D} \sim \mathscr{N} ( \bm{\hat{\mu}}, \bm{\hat{\Sigma}} ).
\end{align}
This conditioning can be done analytically (see \citet[Chp. 2.2]{garnett_bayesoptbook_2023} for formulae for $\bm{\hat{\mu}}$ and $\bm{\hat{\Sigma}}$) and effectively tells the model to assign higher probability to functions that fit the training data well.

The quality of this posterior distribution (in terms of the mean accuracy and variance calibration), however, is highly dependent on the choices made for the functions $m$ and $k$. 
A typical choice for the mean function is $m \equiv 0$, which assumes no prior knowledge of the function being modelled and ensures model predictions from \eqref{eq:posterior} are heavily influenced by the training data. 
The choice of covariance kernel is formed via our prior belief in the expected behaviour of the true function being modelled (e.g. non-periodicity and smoothness).
The covariance function used here is the Mat{\'e}rn-$(1/2)$ (or exponential) kernel
\begin{align*}
    k(\bm{x}_i, \bm{x}_j) = \sigma^2\mathrm{exp}\left(-\frac{\| \bm{x}_i-\bm{x}_j \|_2}{\ell}\right),
\end{align*}
where $\| \cdot \|_2$ denotes the Euclidean distance \citep{pandit2019comparative}.
In addition, the parameters $\ell$ and $\sigma$ define the input length scale (smaller values produce more `wiggly' functions) and the function noise (smaller values lead to lower predictive uncertainty in the function). 
The covariance kernel encodes the relationship between input points and the resulting covariance matrix quantifies how a change in one point influences changes in another across the domain.
The \emph{hyperparameters} $\ell$ and $\sigma$ are tuned (for example, using traditional non-Bayesian optimisation algorithms) to produce the best fit to the training data such that the marginal log-likelihood of the posterior \eqref{eq:posterior} is maximised.

It should be noted that while we have described \emph{scalar} output GPs here, in practice we model each output dimension of $\bm{f}$ using its own scalar GP.
This assumes each output of $\bm{f}$ is uncorrelated (i.e. independent) of one another and means that we require $l+m$ ``stacked'' GPs to model the joint distribution over $\bm{f}$.
More importantly for BO, it is crucial that the surrogate model is relatively cheap to train and evaluate compared to the cost of evaluating $\bm{f}$.

\subsection{The classifier} \label{sec:classifier}
The aim of binary classification is to label each data point in the input space as either a failure ($0$) or a non-failure ($1$). 
The probability that a point $\bm{x}\in\mathscr{U}$ is a non-failure is modelled using a GP over a latent function, which is then transformed via the logistic function\citep{williams1998bayesian}:
\begin{equation} \label{eq:classifier}
    p \left( \mathrm{non{\text-}failure} \ | \ \bm{x} \right) = \frac{1}{1+e^{-f(\bm{x})}} \quad \bm{x} \in \mathscr{U}.
\end{equation}
Using the logistic function transforms the GP from a regression model (as it is used in \cref{sec:gps}) to a classifier by mapping its prediction into a probability that a point belongs to the non-failure class.


The GP model with classification now has a Bernoulli likelihood $p(\mathscr{D}|f)$, making the calculation of the posterior distribution $p(f|\mathscr{D})$ analytically intractable---unlike in \eqref{eq:posterior} where the prior, likelihood, and, therefore, posterior were all Gaussian (see \citet[Chp. 3.4]{williams2006gaussian}). 
To address this, we approximate the posterior using a variational distribution $q(f | \mathscr{D}; \bm{\lambda})$, chosen such that its likelihood $q(\mathscr{D} | f ; \mathbf{\bm{\lambda}})$ is Gaussian---with $\bm{\lambda}$ parameterising the new distribution \citep{tran2015variational}.
The parameters $\bm{\lambda}$ are found by maximising the evidence lower bound
\begin{equation*}
    \mathscr{L}(\bm{\lambda}) \defeq \mathbb{E}_{q(\mathscr{D}|f;\bm{\lambda})} \left[ \log p(\mathscr{D}|f) \right] - \mathrm{KL} \left[ q(f;\bm{\lambda}) \ \| \ p(f) \right].
\end{equation*}

The first term represents the expected log likelihood (observing the training data given the probability distribution over functions), while the second term denotes the (non-negative) Kullback--Leibler divergence between the two distributions.
Clearly if the KL divergence was zero (the distributions were identical), we would be maximising over the original (log) Bernoulli likelihood. 
Once the $\bm{\lambda}$ are found, the GP can be conditioned on the data (as was shown in \cref{sec:gps}) using the new Gaussian likelihood $q(\mathscr{D}|f;\mathbf{\lambda})$. 

In classification, imbalance in the dataset--where one label is significantly more prevalent--can create a poor quality classifier.
This results in the classifier being accurate by simply predicting the majority class, rendering it useless for identifying failure regions.
To combat this, we employ \emph{oversampling}, which randomly duplicates samples in the minority class such that both labels are equally represented in the training dataset \citep{imbalanced2013}. 
As a result, the classifier cannot achieve a high accuracy by simply predicting one class and a higher quality model is produced.

\subsection{The acquisition function} \label{sec:acqf}

Based on knowledge from the trained GP and classifier, the acquisition function provides us with a way to estimate how informative evaluating $\bm{f}$ at a previously unseen point $\bm{x} \in \mathscr{U}$ will be.
Depending on the task at hand, there are many possible choices of acquisition functions, each tailored to specific objectives. 
As mentioned before, the key factor in selecting an appropriate one is that it should be computationally cheap (compared to $\bm{f}$) to evaluate given the surrogate model.

Here, we use the \emph{expected hypervolume improvement} (EHVI) function, which seeks to quantify the expected increase in the hypervolume of $\mathscr{P}_{\bm{f}}$ when adding a new point to the dataset $\mathscr{D}$ \citep{daulton2020differentiable}.
The \emph{hypervolume} HV of $\mathscr{P}_{\bm{f}}$ is defined as the $l$-dimensional integral of the subspace 
\begin{equation*}
   \{ \bm{y} \in \Reals^{l} \mid \exists \bm{p} \in \mathscr{P}_{\bm{f}} \ \text{s.t.} \ \bm{p} \prec \bm{y} \},
\end{equation*}
dominated by $\mathscr{P}_{\bm{f}}$ \citep{yang2019multi}.

EHVI is particularly suited to multi-objective optimisation, as it effectively balances exploration and exploitation by focusing on regions of the search space that are both uncertain and potentially optimal.
The EHVI function $\alpha_{\text{EHVI}} \colon \mathscr{U} \to \Reals$ is given by 
\begin{equation*}
    \alpha_{\text{EHVI}}(\bm{x}) = \mathbb{E}_{\bm{f}} \left[ \text{HV}(\mathscr{P}_{\bm{f}} \cup \{ \bm{f}(\bm{x}) \}) - \text{HV}(\mathscr{P}_{\bm{f}}) \right],
\end{equation*}
where $\mathbb{E}_{\bm{f}}$ is the expectation operator of \eqref{eq:posterior} (with respect to the $l$ objectives, not the constraints).
Recalling that $\bm{f}(\bm{x})$ is a random variable, this function describes how much additional volume in objective space we expect to gain by sampling at a new point $\bm{x}$, relative to the current Pareto front. 
Please refer to \citet{yang2019multi} for a more rigorous treatment of this material.

As mentioned before, we have both constraints on the function $\bm{f}$ and failure regions in the input space.
To this end, we define the \emph{probability of feasibility} as
\begin{equation*}
    g(\bm{x}) = p(\mathrm{non{\text-}failure}|\bm{x}) \ \prod_{i = 1,\ldots,m}p(f_{l+i}(\bm{x})\leq0),
\end{equation*}
where $f_{l+1}$ to $f_{l+m}$ are the GP models of the $m$ constraint functions and $p(\mathrm{non{\text-}failure}|\bm{x})$ is the classification model \eqref{eq:classifier}.
This measures the joint probability that a given point in the input space is feasible (respects all of the constraints) and is not a failure.

Using this probability, we can then define a \emph{constrained} acquisition function (ECHVI):
\begin{equation} \label{eq:echvi}
    \alpha_{\text{ECHVI}}(\bm{x}) = \begin{cases}
        \alpha_{\text{EHVI}}(\bm{x})\cdot g(\bm{x}) & \text{if } g(\bm{x}) \geq \lambda, \\
        0 & \text{otherwise.}
    \end{cases}
\end{equation}
which weights $\alpha_{\text{EHVI}}$ by $g$ and enforces a cut-off threshold\citep{gardner2014bayesian, abdolshah2018expected} (here we use $\lambda = 0.5$).
This ensures that samples with a probability of feasibility less than $\lambda$ are excluded from consideration, while those with probabilities higher are more likely to be selected during the optimisation than those with lower probability (but still above $\lambda$). 
We also note that $\alpha_{\text{EHVI}}$ must be calculated with respect to the hypervolume of the \emph{feasible} Pareto frontier by excluding infeasible points from the dataset.

To find the next most informative sample, we use single-objective (non-Bayesian) optimisation to find the point that maximises $\alpha_{\text{ECHVI}}$:
\begin{align*}
    \bm{x}^* = \underset{\bm{x} \in \mathscr{U}}{\arg\max} \ \alpha_{\text{ECHVI}}(\bm{x}).
\end{align*}
This is done using the L-BFGS-B\citep{byrd1995limited} algorithm which makes use of multiple restarts to avoid local maxima and avoid the discontinuity in $\alpha_{\text{ECHVI}}$.
Once found, $\bm{x}^*$ is evaluated using $\bm{f}$ and added to the dataset $\mathscr{D}$.

\section{The poloidal field coil design problem} \label{sec:design-problem}

The PF coil design problem described here is concerned with identifying the set of PF coil positions that will optimise some aspects of both cost and performance of a STEP-like tokamak, subject to strict design and engineering requirements.
In this section, we will describe the inputs, objectives, and constraints required to formulate the optimisation problem as well as the underlying STEP baseline design and the simulator required to calculate plasma equilibria. 
Throughout, we will be working within a cylindrical coordinate system $(R, \phi, Z)$ which denotes the major radius, the toroidal direction (into the page), and the height, respectively.

\subsection{The STEP baseline design} \label{sec:the-baseline}

\begin{figure}[t]
    \centering
    \includesvg[width=0.4\textwidth, keepaspectratio]{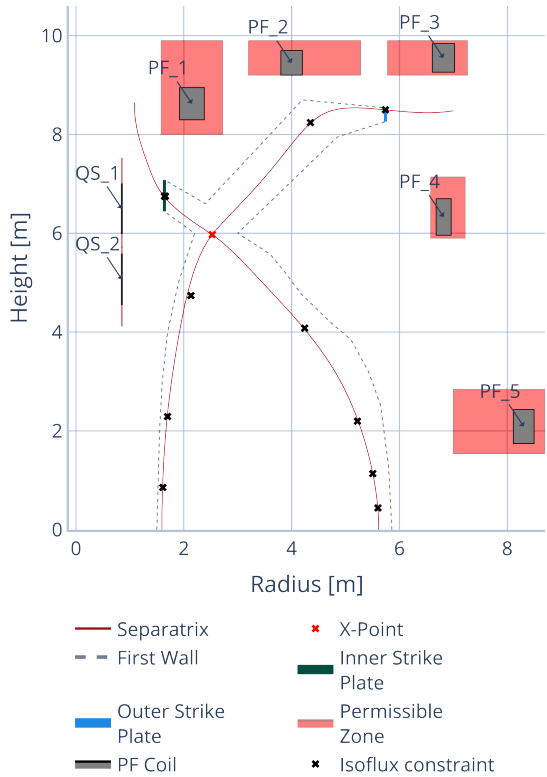}
    \caption{The baseline setup from \citet{hudoba2023magnetic} in the $RZ$-plane (due to vertical symmetry about $Z=0$ only the top half of the tokamak is shown). 
    The separatrix (red) outlines the shape of the plasma core and the divertor legs which hit the inner (green) and outer (blue) strike plates. 
    The initial location of the seven PF coils (grey rectangles) are displayed within their respective permissible zones (red rectangles).
    Note the absence of the central solenoid, which is not used in the flat-top phase of operation shown here. 
    The isoflux constraints (black crosses) define locations at which the separatrix should pass through.}  
    \label{fig:baseline}
\end{figure}

We will be working with the initial PF coil setup and limiter geometry from the baseline STEP design presented by \citet{hudoba2023magnetic}.
The design is shown in \cref{fig:baseline} and the information available to us from the baseline dataset are the:
\begin{itemize}[label=$\circ$] 
    \item Names, centroid positions $(R^c,Z^c)$, and half width/heights $(\mathrm{d}R,\mathrm{d}Z)$ of the PF coils. 
    \item Permissible zones for each PF coil, i.e. the region of the $RZ$\nobreakdash-plane in which each coil can be placed without intersecting the TF coils, diagnostic systems, or other parts of the tokamak.
    \item Limiter contour that confines the plasma equilibrium. This was constructed using the strike plate locations and made to match the geometry illustrated in \citet{Tholerus_2024}.
    \item Strike plate locations, i.e. segments of the limiter in the inner and outer divertor where the legs of the plasma separatrix will strike.
    \item Separatrix of the plasma equilibrium, the X-points, and the strikepoints. 
    \item Plasma pressure and toroidal magnetic field profiles required to solve for the equilibrium.
\end{itemize}

\subsection{Optimisation problem} \label{sec:problem-formulation}

The mathematical formulation of the PF coil optimisation problem requires stacked scalar inputs, objectives, and constraints so that we can map a vector of PF coil positions to a vector of objective/constraint values.

\subsubsection{Input space}
As can be seen in \cref{fig:baseline}, there are seven up-down symmetric (around $Z=0$) PF coil circuits each with their own $(R^c,Z^c)$ centroid coordinate that is allowed to move freely such that no part of the coil leaves the permissible zone.
The exceptions are the two quasi-solenoid (\emph{QS}) coils, positioned above (and below) the central solenoid for magnetic shaping in the inner divertor, which are only able to move vertically.
This results in a twelve dimensional input space for the optimisation problem: five pairs of $(R^c,Z^c)$ coordinates for the \emph{PF} coils and one $Z^c$ coordinate for each of the two \emph{QS} coils. 
We normalise each of the coordinates with respect to their own permissible zones so that we can work with the unit hypercube $[0,1]^{12}$ as our input space.
A more detailed explanation of the normalisation process can be found in \cref{app:coil-parameterisation}.

\subsubsection{Objectives}
In this problem, we consider two scalar objective functions that we wish to optimise with multi-objective BO---though we should note that nothing prevents us from adding more objectives.

\begin{figure}[t!]
    \centering
    \includesvg[width=0.4\textwidth, keepaspectratio]{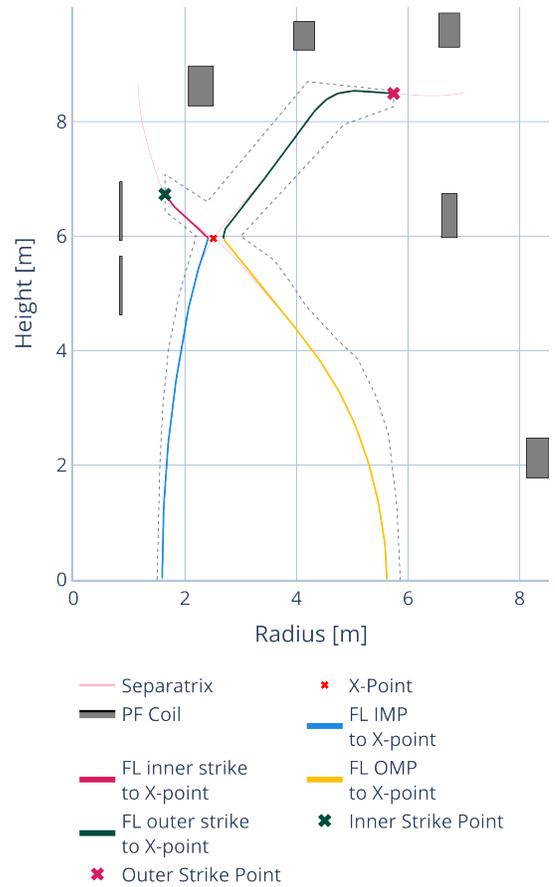}
    \caption{The baseline setup from \cref{fig:baseline}, this time marked with objective and constraint quantities.
    Shown are the separatrix (light pink), the upper X-point (red cross), and the inner (green star) and outer (red star) strikepoints. 
    Also shown are the flux surfaces traced out to calculate the ICL (blue plus red lines) and OCL (yellow plus green lines).}
    \label{fig:metrics}
\end{figure}

The first objective is to \emph{minimise} the volumetric sum of the PF coils. This is important as smaller coils require less physical material and therefore weigh less, making the fabrication, transportation, and installation process less arduous and costly.
Recall, each PF coil is modelled as a rectangle in the $RZ$\nobreakdash-plane and as an annulus in the $R\phi$\nobreakdash-plane.
The total volume of the seven PF coils (upper and lower components inclusive) can therefore be defined as 
\begin{equation} \label{eq:volume}
    V =  8 \pi \sum_{i=1}^{14} R^c_i \mathrm{d}R_i \mathrm{d}Z_i.
\end{equation}
where the coil index $i = 1,\ldots,7$ for upper and $i = 8,\ldots,14$ for lower coils.
It is important to note that the PF coils do not change shape in the poloidal plane (i.e. cross-sectional areas are fixed, equal to $4 \mathrm{d}R \mathrm{d}Z$) and so the volume of each coil will change only when its central radial position $R^c$ changes.

The second objective is to \emph{maximise} the average of the inner connection length (ICL) and the outer connection length (OCL). 
The connection length is the distance traced out by a helical (i.e. moving both poloidally and toroidally) magnetic field line that starts at the inner (outer) edge of the last closed flux surface (LCFS) at the midplane and ends at the inner (outer) strikepoint \citep{doyle2021magnetic, albanese2023three}.
We refer to these inner and outer midplane points as the IMP and OMP, respectively.
Larger connection lengths ensure that hot plasma leaving the core edge region will travel a farther distance and therefore cool to more acceptable temperatures before hitting the strike plate.
This is an important aspect of managing heat loads in the divertor region. 

The ICL and OCL are calculated by carrying out an integral over the length of the magnetic field line of interest \citep{kos2019smiter}. 
Tracing the magnetic field line requires the solution of a system of ordinary differential equations (ODEs) for the position vector of a point along the field line trajectory $\bm{r}(\ell)$:
\begin{align} \label{eq:connec_length}
    \frac{\mathrm{d}\bm{r}(\ell)}{\mathrm{d} \ell} = 
    \bm{b} \left( \bm{r}(\ell) \right),
    \quad \ell \in [0,L],
\end{align}
where $\bm{b}$ is a unit vector along the direction of the magnetic field $\bm{B}$ in cylindrical coordinates:
\begin{align*}
    \bm{B} \left( \bm{r}(\ell) \right) =
    \begin{cases} 
        -\frac{1}{R} \frac{\partial \psi(R,Z)}{\partial Z} \\[6pt]
        \frac{F(\psi(R,Z))}{R}. \\[6pt]
        \frac{1}{R} \frac{\partial \psi(R,Z)}{\partial R}
    \end{cases}
\end{align*}


Here $\psi(R,Z)$ denotes the scalar poloidal magnetic flux and $F$ the toroidal magnetic field profile.
To solve this problem, we use a fourth-order Runge-Kutta method (with appropriately chosen step size $\Delta \ell$) and integrate until a terminating condition is met (see next paragraph), recording the value of $L$ obtained (i.e.\ the total number of steps $\Delta \ell$).

In practice, however, the calculation is split into two stages. 
The first stage involves integrating from an initial position $\bm{r}(0)$ which is selected to be $3$mm radially outside the IMP/OMP and ending at some very small distance away from the upper X-point.
The second stage integrates between the inner/outer strikepoint and the point close to the upper X-point. 
The length of these individual sections is then combined to return the final connection length.
The two flux surfaces traced out when calculating the ICL and OCL are visualised in \cref{fig:metrics}.
These techniques reduce the likelihood of the integrator getting stuck at the exact X-point, travelling around the LCFS (instead of going up into a divertor), and from travelling into the wrong divertor.

\subsubsection{Constraints}
In addition to the objectives, we also have a number of design and engineering constraints that need to be satisfied so that each PF coilset considered in the BO loop produces an equilibrium with key targets that are similar to the baseline equilibrium and does not violate coil current limits.
The bounds of the constraints are summarised in \cref{tab:constraints}.
\begin{table}[t]
    \centering
    \begin{tabular}{|l|l|}
        \hline
        \textbf{Constraint} & \textbf{Bound [unit]} \\
        \hline\hline
        LCFS area ratio & $\leq 0.012$ \\
        \hline
        Outer strike distance & $\leq 0.14$ [m] \\
        \hline
        Inner strike distance & $\leq 0.32$ [m] \\
        \hline
        X-point distance & $\leq 0.01$ [m] \\
        \hline
        Maximum current density & $\leq 100$ [$\mathrm{MA/m^2}$] \\
        \hline
    \end{tabular}
    \caption{Constraint bounds enforced on the equilibria generated by FreeGS for a particular PF coilset.}
    \label{tab:constraints}
\end{table}

The first constraint is on the shape of the LCFS, which is defined as the contour of $(R,Z)$ points that pass through the X-point closest to the magnetic axis---recall \cref{fig:baseline}. 
We denote this region of the $RZ$\nobreakdash-plane as $\Omega_p$ and quantify the difference between two different regions using
\begin{align*}
    \eta \big( \Omega_{p}^{1}, \Omega_{p}^{2} \big) \defeq \frac{\big| \Omega_p^{1} \cup  \Omega_p^{2} \big| - \big| \Omega_p^{1} \cap \Omega_p^{2} \big|}{ \big| \Omega_p^{1} \big| + \big| \Omega_p^{2} \big|} \in [0,1],
\end{align*}
where $| \cdot |$ denotes the cross-sectional area of a region in the poloidal plane \citep{bardsley2024decoupled}.
This parameter quantifies the ratio of the total non-overlapping areas and the sum of the two areas. 
Placing an upper limit on this ratio enables us to constrain the LCFS shape of the new equilibrium ($\Omega^2_p$) to be similar to that of the baseline ($\Omega^1_p$). 
This helps to ensure the new equilibrium has similar core performance to the baseline.

The second and third constraints place an upper limit on the distance between the strikepoints (i.e. where the separatrix first intersects the limiter geometry at some location) and the centre of the strike plates.
In rare cases, an equilibrium may have a separatrix that intersects the limiter multiple times on the same plate, hence we need to account for that. 
The bound is half of the length of the strike plate, with one constraint on each of the inner and outer strike plates.

The fourth constraint will place an upper limit on the distance between the two X-point locations when mirrored about $Z=0$. 
This distance should be minimal in a double-null plasma scenario as considered here. 
See \cref{fig:metrics} for the strikepoint and X-point locations. 

The final constraint ensures the maximum current density 
\begin{equation*}
    J_{\text{max}} = \frac{1}{4}\max_{i\in[1...7]} \frac{I_i}{\mathrm{d}R_i \mathrm{d}Z_i},
\end{equation*}
in the PF coilset remains below the engineering limit defined in \citet{nasr2024}. 
Here, $I_i$ denotes the coil current and the denominator is the coil area. 
This limits stresses in the PF coil structures and helps avoid quench events---a sudden loss of superconductivity which can damage the coils \citep{coatanea2015electromagnetic}. 

\subsubsection{Equilibrium simulator}
\label{sec:simulator}
In order to calculate the aforementioned objective functions and evaluate whether or not the constraints have been met, we need a simulator that is able to generate a plasma equilibrium using the STEP baseline design and a given PF coilset.
For this we use \emph{FreeGS}, a free-boundary static inverse equilibrium solver\cite{freegs}. 
FreeGS will return a plasma equilibrium (in terms of the poloidal flux) and the PF coil currents required to generate it.
It uses an optimisation routine to identify the coil currents with respect to some constraints on the chosen plasma shape and a Picard iteration scheme to solve the free-boundary Grad-Shafranov problem (see \citet{song2024} and \citet{pentland2024}). 
The required inputs to solve the equilibrium problem are:
\begin{itemize}
    \item The STEP baseline parameters and a PF coilset (permissible zones not required).
    \item Two X-point locations, one at $(R^X,Z^X)$ and the other mirrored at  $(R^X,-Z^X)$ (as we required an up-down symmetric double-null configuration like the baseline equilibrium).
    \item 23 isoflux constraints that link poloidal flux values on the core plasma boundary to the X-points and the divertor regions (i.e. constraints that ensure the poloidal flux $\psi(R,Z)$ at two different locations $(R_1,Z_1)$ and $(R_2,Z_2)$ are the same).

\end{itemize}
Given these inputs, FreeGS will return the coil currents in the PF coils required to generate an equilibrium that (closely) matches the one provided in the baseline.
From this equilibrium, we can then calculate the values of the objectives and the constraints. 
From time to time, however, the simulator may fail to converge on a physically ``valid'' equilibrium, returning spurious objective and constraint values.
This could be for a number of reasons such as solver instability or a physically incompatible PF coilset. 
This requires care and will be discussed in the next section. 

\subsubsection{Failure regions} \label{sec:failure}

\begin{figure}[t]
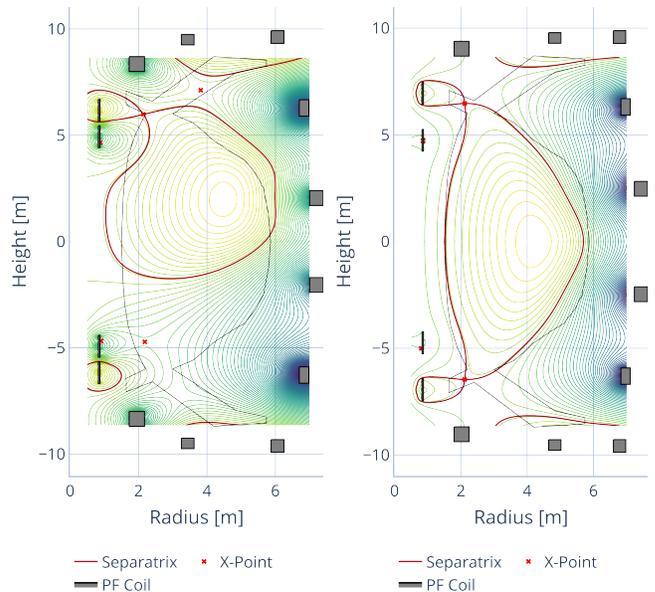

    \centering
    \begin{subfigure}[t]{0.49\linewidth}
        \includesvg[width=0.99\textwidth, keepaspectratio]{figures/bad_eqm_1.svg} 
        \label{fig:invalid-equilibria-xpoint}
    \end{subfigure}
    \begin{subfigure}[t]{0.49\linewidth}
        \includesvg[width=0.99\textwidth, keepaspectratio]{figures/bad_eqm_2.svg} 
        \label{fig:invalid-equilibria-bad}
    \end{subfigure}
    \caption{Two examples of invalid equilibria returned by ``converged'' FreeGS simulations with different PF coilsets.}
    \label{fig:invalid-equilibria}
\end{figure}

The equilibrium simulator will either (in rare cases) fail to converge or stop once the relative difference between the poloidal flux at successive iterations is below some tolerance threshold (returning an equilibrium solution). 
This stopping criteria does not, however, consider the physical validity of the equilibrium identified.
In some cases, non-physical equilibria that do not satisfy the X-point and isoflux constraints may be returned. 
In other cases, we may have an equilibrium for which we either cannot calculate the objectives/constraints or which return spurious objective/constraint values.
The regions of the input space for which non-physical (\emph{invalid}) equilibria are returned (or if the simulator outright fails) will be referred to as \emph{failure regions}.

In \cref{fig:invalid-equilibria}, we illustrate two cases of invalid equilibria returned by FreeGS. 
The left panel shows a single-null equilibrium with the LCFS intersecting the limiter in the core region. The right panel depicts an equilibrium in which both X-points have formed inside the divertor regions, far from the desired locations in the core, resulting in the LCFS again intersecting the limiter. 
This is problematic when calculating the ICL, as this calculation assumes the LCFS does not intersect the limiter geometry until hitting the strikeplate, resulting in an unfeasibly small ICL value. 
Calculations for both the ICL/OCL and the strike distances are spurious in this case.

To mitigate these issues, we can classify (recall \cref{sec:classifier}) whether an equilibrium is valid by checking the following conditions:
\begin{enumerate}
    \item the X-points $({R^X,Z^X})$ and $({R^X,-Z^X})$ must be to within $10$cm of the limiter boundary with $R^X \in [2.2, 3]$.
    \item the LCFS does not intersect the limiter\footnote{There is now an extension to FreeGS---\emph{FreeGSNKE} \citep{amorisco2024}---which ensures the core remains within the limiter. Use of this equilibrium solver would remove the need for this constraint, however, it was released following the completion of this work.}.
\end{enumerate}
By actively avoiding sampling the PF coilsets where the simulator fails or produces such invalid equilibria (via the classifier), we can avoid wasting computational resources on solutions that do not provide any useful information to the BO loop.

\section{Numerical experiments} \label{sec:results}

In this section, we will perform the design optimisation of the PF coil set problem.
The first experiment will use multi-objective BO to find several Pareto optimal PF coilsets that respect the engineering and design constraints in \cref{sec:problem-formulation}.
We analyse two of the Pareto optimal solutions in more detail, highlighting how the BO explores the solution space while respecting the trade-off between the objective functions. 
To further illustrate the data efficiency of the BO, we compare its performance against two other optimisation methods (simple Sobol sampling and a genetic algorithm) when using both identical and larger computational budgets. 

To generate these results, we use the Trieste \citep{trieste2023, Berkeley_Trieste_2025} package which  provides the software implementations for Sobol sampling, acquisition functions, and Gaussian processes (via GPflow \citep{GPflow2017}). 
Pygmo2 \citep{Biscani2020} provides the genetic algorithm which we will use for benchmarking.
When evaluating the Sobol samples with FreeGS\citep{freegs}, we make use of the CSD3 HPC cluster (see Acknowledgements) and the Simvue platform \citep{Lahiff_Simvue_2024} to monitor simulation progress and store the objective/constraint data. 

\subsection{Stand-alone BO}
\label{sec:bo-optimisation-of-pf}
In this experiment, we will limit ourselves to $128$ evaluations of $\bm{f}$: $64$ Sobol samples to build the initial dataset and $64$ sequential BO samples to intelligently explore the objective space and identify feasible optimal points. 

In \cref{tab:convergence}, we display the proportions of each sampling set that result in feasible, infeasible (violating one or more constraints), and failed (invalid) PF coilsets. 
We can see that only $10\%$ of the Sobol samples provide feasible designs and that once the BO loop begins running, we accumulate a much larger proportion of feasible designs with fewer failures.
This shows that the GPs can accurately model the constraint responses and the acquisition function uses this to propose feasible samples.

The drop in failure region samples likely results from a combination of explicit classifier intervention and the scarcity of optimal samples near these regions, making them less likely to be chosen by the acquisition function. 
The classifier has a precision of $0.82$, meaning $82\%$ of the area included in the acquisition maximisation (by zeroing the ECHVI in these regions) is indeed non-failing, reducing waste of computational resources by potentially sampling failing points. 
A recall of $0.86$ shows that only $14\%$ of the non-failure region is incorrectly avoided by the classifier; it is more important that this percentage is low because Pareto optimal solutions could exist here but would be missed.

\begin{table}[b]
    \centering
    \begin{tabularx}{0.48\textwidth}{|X|c|c|c||c|}
       \hline
       \textbf{Method} & \textbf{Failure} & \textbf{Infeasible} & \textbf{Feasible} & \textbf{Total} \\
       \hline\hline
        Sobol & $23\:(36\%)$ & $31\:(48\%)$ & $10\:(16\%)$ & $64$
        \\\hline
        BO & $4\:(6\%)$ & $37\:(58\%)$ & $23\:(36\%)$ & $64$
        \\\hline
    \end{tabularx}
    \caption{The number (and percentage) of samples from each sampling method that lie in failure, infeasible, or feasible regions.}
    \label{tab:convergence}
\end{table}

\begin{figure}[t]
    \centering
    \includesvg[width=0.5\textwidth, keepaspectratio]{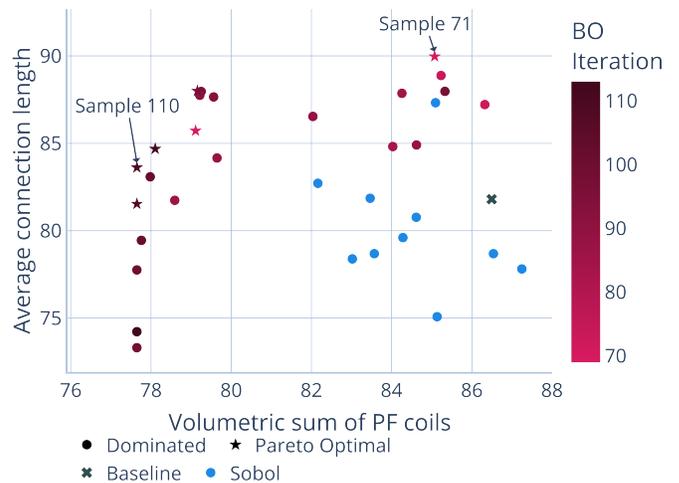}
    \caption{Feasible solutions in the objective space for the $64$ Sobol + $64$ BO experiment.
    Shown are the baseline (grey cross), Sobol solutions (blue) and the BO solutions (light pink to red). 
    Pareto optimal solutions are denoted with a star and dominated solutions with a circle, with light pink to red indicating successive BO iterations.
    Two of the BO samples, $71$ and $110$, are highlighted for further analysis.}
    \label{fig:pareto-frontier}
\end{figure}

\begin{figure}[b]
    \centering
    \includesvg[width=0.49\textwidth, keepaspectratio]{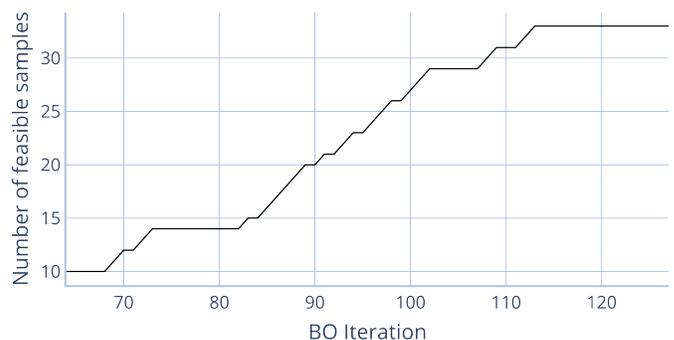}
    \caption{The total number of feasible BO samples at the end of each BO iteration.}
    \label{fig:feasible-increase}
\end{figure}

The BO loop successfully identifies the Pareto front (see \cref{fig:pareto-frontier}), capturing the trade-off between the volumetric sum \eqref{eq:volume} and the average connection length (recall \eqref{eq:connec_length}).
Generally, the latter BO iterations produce samples that dominate earlier samples, highlighting how BO learns from new data and exploits its new understanding of the functions to produce higher-quality samples.

During the initial BO iterations, the data is sparse, resulting in highly uncertain and inaccurate GPs. 
As a result, the exact location of the feasible regions is unclear and, as it explores for the first $20$ iterations, the BO produces few feasible samples---see \cref{fig:feasible-increase}. 
Following this initial exploration, $30$ iterations of exploitation takes place, where BO reliably produces feasible samples (those seen in \cref{fig:pareto-frontier}). 
The final $20$ iterations yield no feasible samples, indicating a return to exploration and suggesting there is little scope to find feasible optimal solutions around the current Pareto optimal points. 

\begin{table}
    \centering
    \begin{tabularx}{0.49\textwidth}{|X|c|c|c|}
        \hline
         \textbf{Objective/constraint} & \textbf{Baseline} & \textbf{Sample $\bm{110}$} & \textbf{Sample $\bm{71}$} \\
         \hline\hline
         $V$ [$\mathrm{m}^3$] & $86.5$  & $77.6$ & $85.1$ \\
         \hline
         ICL [$\mathrm{m}$] & $96.5$ & $96.9$ & $108$ \\
         \hline
         OCL [$\mathrm{m}$] & $67.1$ & $70.3$  & $71.8$ \\
         \hline\hline
         LCFS area ratio & $0.0108$ & $0.0117$ & $0.0103$ \\
         \hline
         Outer strike distance [$\mathrm{m}$] & $0.0934$ & $0.0957$ & $0.0962$ \\
         \hline
         Inner strike distance [$\mathrm{m}$] & $0.0297$ & $0.0233$ & $0.0305$ \\
         \hline
         X-point distance [$\mathrm{m}$] & $0.000374$ & $0.00394$ & $0.000919$\\
         \hline
         $J_{\text{max}}$ [$\mathrm{MA}/\mathrm{m}^2$] & $71.4$ & $82.4$ & $98.8$ \\
         \hline\hline
         Inner $\langle B_p \rangle$ [$\mathrm{T}$] & $0.414$ & $0.413$ & $0.373$ \\
         \hline
         Outer $\langle B_p \rangle$ [$\mathrm{T}$] & $0.552$ & $0.537$ & $0.530$ \\
         \hline
    \end{tabularx}
    \caption{The objectives and constraint values (to three significant figures) for the baseline and two of the Pareto optimal solutions shown in \cref{fig:pareto_equilibria}.
    Also shown are the inner and outer line-averaged poloidal magnetic field readings.
    }
    \label{tab:key-quantities}
\end{table}

\begin{figure}[t]
    \centering
    \includesvg[width=0.49\textwidth, keepaspectratio]{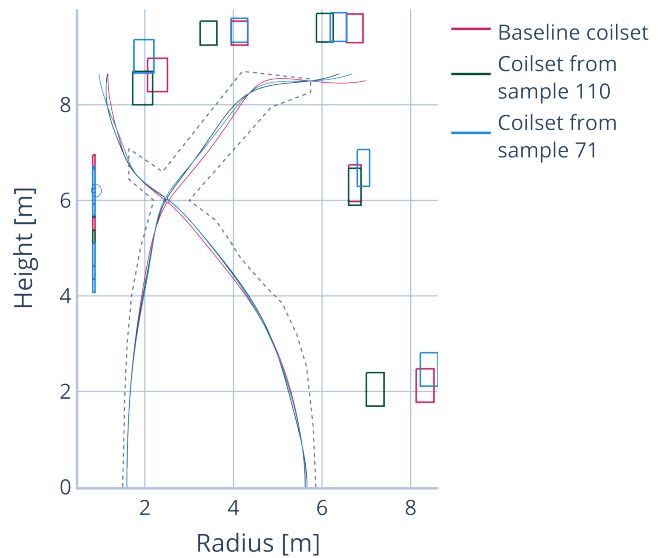}
    \caption{The PF coilsets and corresponding separatrices for the baseline (pink) and two of the Pareto optimal equilibria: sample $71$ (blue) and $110$ (green) of the simulator (corresponding to samples from the $7$th and $46$th BO iteration, respectively).}
    \label{fig:pareto_equilibria}
\end{figure}

In \cref{tab:key-quantities}, we display the objective/constraint values obtained from the baseline and two of the Pareto optimal PF coil sets shown in \cref{fig:pareto_equilibria}, with both samples obtained during the BO iterations.
The $71$st sample yields the highest average connection length while the $110$th sample has the joint lowest volumetric sum (of these tied samples, it has the higher connection length).
The first three rows of the table show the objective quantities for both samples while the intermediate five rows show the constraint values. 

The objective values of the baseline (and its location in \cref{fig:pareto-frontier}) relative to the Pareto optimal samples demonstrate that BO is able to improve the PF coilset design significantly. 
At a minimum, BO has yielded a reduction of $V$ by $1.4\mathrm{m}^3$ and an increase in average connection length of $1.8\mathrm{m}$ over the baseline; this is not insignificant considering the low computational budget to achieve these gains.
The constraints show that samples $110$ and $71$ are close to the constraint bounds for the LCFS shape and maximum coil current density, respectively. 
This could indicate that further optimisation of these samples (and other Pareto optimal samples) is not possible without violating the constraints, hence the lack of feasible samples in the final BO iterations.

It is clear from \cref{fig:pareto_equilibria} that the $110$th sample has a smaller volumetric sum because PF coils 2, 3, 4, and 5 are closer to the device centreline ($R = 0$). However, the difference in average connection length is less obvious because the separatrices look (qualitatively, at least) very similar. The lower connection length in sample $110$ results from a higher poloidal field, causing particles travelling from the midplane into the divertors to move faster, decreasing the number of times (and thus the distance) they orbit the tokamak toroidally. This can be seen in the final two rows of \cref{fig:pareto_equilibria}, which shows the line-averaged poloidal magnetic field $\langle B_p \rangle$ along the inner and outer connection length field lines, respectively.

\subsection{Comparison between BO, Sobol sampling, and a genetic algorithm}

Next, we compare the BO scheme against two other multi-objective optimisation methods.
The first method we compare against uses quasi-Monte Carlo sampling via the Sobol method, essentially relying on random chance to sample feasible and optimal coilsets. 
The second method will use a genetic algorithm, specifically the `Non-Dominated Sorting Genetic Algorithm II' (NSGA-II) \citep{deb2002fast}. 
NSGA-II, like all genetic algorithms, operates on the principle that combining the inputs of well-performing individuals within a population can produce offspring (new samples) that inherit characteristics from its parents and therefore may perform similarly or better. 
Starting with an initial population (in our case, Sobol samples), the algorithm generates additional samples through iterative recombination and mutation (applying slight random changes to the inputs), therefore introducing variation to explore the solution space \citep{verma2021comprehensive}.
Constraints are handled by penalising the objectives according to the number of violated constraints \citep{kuri2002penalty}. 
Similarly, the failure region is handled by returning large constants for the objectives, artificially making the sample appear very non-optimal.

We run six additional experiments with each of these methods, the results of which are presented in \cref{tab:experiments-hypervolume}.
The first three (II, III and IV) use the same computational budget as the BO experiment (I) from \cref{sec:bo-optimisation-of-pf}, while the final three (V, VI and VII) have a budget that is $8 \times$ larger. 

Four of the experiments contain results for the NSGA-II genetic algorithm. 
For each budget, we include an NSGA-II experiment with the same initial population ($64$ samples) as Experiment I (experiments III and VI) and another with a lower initial population (experiments IV and VII) but with the ability to run over more generations. 
In both cases, the experiments using a lower initial population size (IV and VII) outperform their counterparts with larger initial population sizes. 
This is because they are able to explore the solution space more widely (using more iterations), thus generating more feasible solutions and making incremental progress each generation towards the Pareto frontier.
This can be seen by the higher percentage of feasible solutions sampled by these experiments compared to the others. 
For the remaining analysis, we will compare only these best performing NSGA-II experiments (IV and VII) against the Sobol sampling and BO.

\begin{figure}[t!]
    \centering
    \includesvg[width=0.48\textwidth, keepaspectratio]{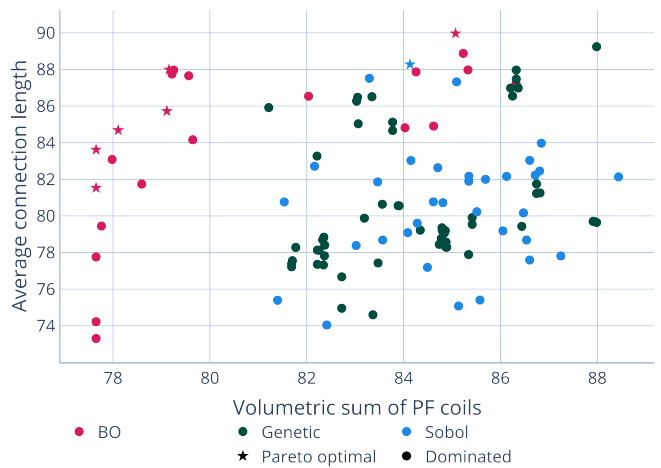}
    \caption{Feasible solutions in the objective space for experiments I (red), II (blue), and IV (green) in \cref{tab:experiments-hypervolume}.
    Pareto optimal solutions are denoted with a star and dominated solutions with a circle.
    }
    \label{fig:concatenated-experiment-64}
\end{figure}

From the results, we can see that experiment I produces better samples than both II and IV with a hypervolume at least $20\%$ larger. 
Recall that a larger hypervolume indicates a feasible objective space with better trade-offs that are further from the (anti-optimal) reference point---the feasible solutions are shown in the objective space in \cref{fig:concatenated-experiment-64}.
The genetic algorithm finds the most feasible samples, outperforming BO by nearly $2\times$. However, the hypervolume of experiment IV indicates few of these feasible samples offer any improvement over even quasi-random samples. 
This illustrates how the BO performs significantly better than the Sobol sampling and the genetic algorithm at finding Pareto optimal PF coilsets. 
Genetic algorithms find the most feasible PF coilsets, however, all of the samples are of significantly lower quality than those from BO.

\begin{table}[t!]
    \centering
    \begin{tabular}{|c|c|c|c|}
        \hline
        & \textbf{Experiment} & $\mathbf{HV}$ & \textbf{Feasible} \\
        \hline\hline
        I & 64 Sobol + 64 BO & $1990.9$ & $26\%$
        \\\hline\hline
        II & 128 Sobol & $1626.6$ & $18\%$
        \\\hline
        III & 64 Sobol + 64 NSGA-II & $1544.2$ & $16\%$
        \\\hline
        IV & 8 Sobol + 120 NSGA-II & $1659.2$ & $49\%$
        \\\hline\hline
        V & 1024 Sobol & $1711.5$ & $13\%$
        \\\hline
        VI & 64 Sobol + 960 NSGA-II & $1713.6$ & $23\%$
        \\\hline
        VII & 32 Sobol + 992 NSGA-II & $1820.3$ & $64\%$
        \\\hline
    \end{tabular}
    \caption{The hypervolume of the feasible solution set and the percentage of total samples taken that were feasible for each of the experiments run.
    Here, we compare the $64$ Sobol + $64$ BO experiment from \cref{sec:bo-optimisation-of-pf} with pure Sobol sampling and the NSGA-II algorithm, each with the same number of samples (128).
    We also display Sobol and NSGA-II experiments that use $8\times$ the number of samples (1024).
    All hypervolumes are calculated with respect to the same reference point.}
    \label{tab:experiments-hypervolume}
\end{table}

\begin{figure}[t!]
    \centering
    \includesvg[width=0.48\textwidth, keepaspectratio]{figures/concatenated-experiments-1024.svg}
    \caption{Feasible solutions in the objective space for experiments I (red), V (blue), and VII (green) in \cref{tab:experiments-hypervolume}. 
    Pareto optimal solutions are denoted with a star and dominated solutions with a circle.
    }
    \label{fig:concatenated-experiment-1024}
\end{figure}

BO continues to outperform Sobol sampling and the genetic algorithm even when we increase their computational budgets to $1024$ samples.
While the hypervolume returned in experiments V and VII are larger compared to those in II and IV (as expected), they still cannot reach the level achieved by the BO (with $1/8$th of the data).
In \cref{fig:concatenated-experiment-1024}, we again see the majority of Pareto optimal samples coming from the BO with a few being found by the genetic algorithm, with BO finding the best samples in each objective (the samples that optimise the marginals of the objective space). 
The vast majority of samples taken by the alternative methods are, however, dominated by others from the BO.
Again, genetic algorithms find the most feasible samples, however, they form a front that underperforms that of BO, particularly in the volumetric sum. 

\section{Discussion and outlook} \label{sec:conclusion}
In this paper, we have demonstrated that BO can successfully identify a set of Pareto optimal PF coilsets in a spherical tokamak. 
Using underlying probabilistic models, it learns the trade-off between the volume of the PF coilset (i.e. the financial cost) and the average connection length produced by the corresponding equilibrium state, simultaneously respecting several physical plasma and engineering constraints. 
Compared to some existing optimisation methods, quasi-Monte Carlo (Sobol) sampling and a genetic algorithm (NSGA-II), BO identifies better solutions while using a significantly smaller computational budget, highlighting its effectiveness and data efficiency.
Overall, the successful application of BO to a complex tokamak design problem should reinforce its suitability for future fusion power plant design challenges, particularly given the increasing reliance on high-fidelity, high-runtime HPC codes where data efficiency is critical.

The relatively poor performance of the Sobol sampling is expected and can likely be attributed to its sparse quasi-uniform coverage of the sample space.
While uniform coverage is good for exploring high dimensional spaces and training emulators (such as the one in our BO loop), the Sobol scheme lacks the ability to hone in on more desirable regions given it is forced to sample inputs within uniformly-spaced partitions of the space. 
NSGA-II outperforms Sobol sampling, especially when both are afforded moderately high computational budgets, however, has an underwhelming performance against BO. 
While it excels at finding feasible samples, NSGA-II fails to find samples dominant over BO, even with a significantly higher computational budget. 
This is likely because the genetic algorithm favours sampling feasible points instead of exploring towards the feasible boundary and potentially finding a more optimal sample---the cost of infeasibility does not outweigh the reward of slight improvements in the objectives. 
It is possible more advanced treatments of the constraints \citep{long2014constraint} would improve the genetic algorithm's performance and allow it to explore closer to the feasible boundary, however, that is beyond the scope of this work.

To increase the applicability and extend this BO framework to ongoing and future PF coil design projects, a number of avenues of future work can be considered.
For example, incorporating additional objectives and constraints should be a trivial task and could be used to help find coilsets that further improve performance. 
For example, one could try to maximise flux expansion to improve divertor performance or include PF coil shaping (in the input space) to try to extract further financial cost savings.
While this would increase the dimensionality of the BO problem, the current framework can be readily adapted to support this via dimensionality reduction techniques \citep{MohitBO_Survey2021, Constantine_2014, grosnit2021highdimensionalbayesianoptimisationvariational}.

\section*{Data Availability Statement}
The data that support the findings of this study are available from the corresponding author upon reasonable request.

\section*{Acknowledgements}
The authors would like to thank Agnieszka Hudoba for providing the baseline STEP data files and Theodore Brown for discussions around BO. 

This work was funded by the EPSRC Energy Programme [grant number EP/W006839/1]. To obtain further information, please contact \href{mailto:PublicationsManager@ukaea.uk}{PublicationsManager@ukaea.uk}.
For the purpose of open access, the author(s) has applied a Creative Commons Attribution (CC BY) licence to any Author Accepted Manuscript version arising.

This work was performed using resources provided by the Cambridge Service for Data Driven Discovery (CSD3), operated by the University of Cambridge Research Computing Service (\url{www.csd3.cam.ac.uk}).
These resources were provided by Dell EMC and Intel using Tier-2 funding from the Engineering and Physical Sciences Research Council (capital grant EP/T022159/1) and DiRAC funding from the Science and Technology Facilities Council (\url{www.dirac.ac.uk}).

\appendix

\section{Coil location normalisation}
\label{app:coil-parameterisation}
\begin{figure}[t]
    \centering
    \includesvg[width=0.49\textwidth, keepaspectratio]{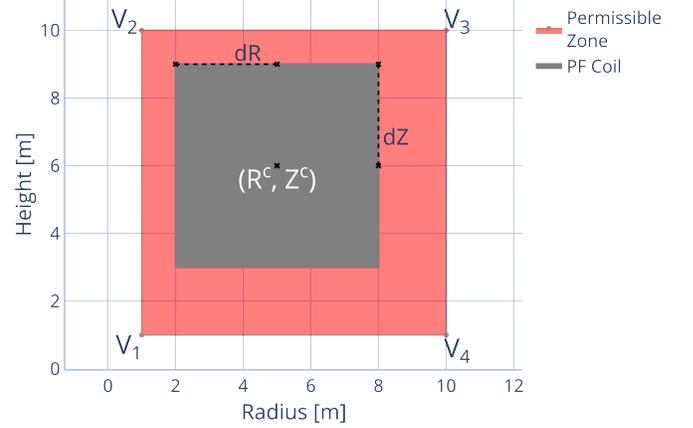}
    \caption{Illustration of a rectangular PF (dark grey), with its centroid and half-width/height marked, and its permissible zone (red), with corner vertices marked. }
    \label{fig:coil-parameterisation}
\end{figure}

Here, we outline how to normalise the centroid coordinates of each PF coil with respect to its permissible zone. 
First, define the lower left and upper right corners of each permissible zone as $\bm{V}_1 = (R_{\text{min}},Z_{\text{min}})$ and $\bm{V}_3 = (R_{\text{max}},Z_{\text{max}})$, respectively.
Given each coil must entirely reside within its permissible zone, we know that the centroid must remain within a half-thickness of the permissible zone:
\begin{align*}
    (R^c, Z^c) \in \left[ R_{\text{min}} + \mathrm{d}R, R_{\text{max}} - \mathrm{d}R \right] \ \times \\
    \left[ Z_{\text{min}} + \mathrm{d}Z, Z_{\text{max}} - \mathrm{d}Z \right].
\end{align*}
We can then obtain the normalised centroid coordinates (with respect to the permissible zone) by defining
\begin{align*}
    \widetilde{R^c} = \frac{R^c - (R_{\text{min}} + \mathrm{d}R)}{(R_{\text{max}} - \mathrm{d}R) - (R_{\text{min}} + \mathrm{d}R)} \in [0,1], \\
    \widetilde{Z^c} = \frac{Z^c - (Z_{\text{min}} + \mathrm{d}Z)}{(Z_{\text{max}} - \mathrm{d}Z) - (Z_{\text{min}} + \mathrm{d}Z)} \in [0,1]. \\
\end{align*}
An illustration of a PF coil and its permissible zone are shown in \cref{fig:coil-parameterisation}. 

\section*{References}
\nocite{*}
\bibliography{aipsamp}

\end{document}